\documentclass[conference]{IEEEtran}

\newcommand{\tabincell}[2]{\begin{tabular}{@{}#1@{}}#2\end{tabular}}

\usepackage[letterpaper, left=1.01in, right=1.01in, bottom=1.01in, top=0.76in]{geometry}
\usepackage{amsfonts}
\usepackage{array}
\usepackage{amsmath}
\usepackage{float}
\usepackage{indentfirst}
\usepackage{amssymb}
\usepackage{epsfig,latexsym}
\usepackage{wrapfig}
\usepackage[justification=normal]{caption}
\usepackage{times}
\usepackage{subfigure}
\usepackage{graphicx}
\usepackage{cite}
\usepackage{subfigure}
\usepackage{bm}
\usepackage{multirow}
\usepackage{algorithm} 
\usepackage{algorithmic} 
\usepackage{multirow}
\usepackage{color}
\usepackage{caption}

\ifCLASSINFOpdf
  \else
  \fi

\begin{document}

\title{Energy-Efficient Adaptive Transmission in Machine Type Communications with Delay-Outage Constraints}

\author{\IEEEauthorblockN{Linlin Zou\IEEEauthorrefmark{1}, Yanzhao Hou\IEEEauthorrefmark{1}\IEEEauthorrefmark{2}, Xiaofeng Tao\IEEEauthorrefmark{1}, Qimei Cui\IEEEauthorrefmark{1}, Xueqing Huang\IEEEauthorrefmark{3}}
\IEEEauthorblockA{\IEEEauthorrefmark{1}National Engineering Lab for Mobile Network Technologies, }
\IEEEauthorblockA{Beijing University of Posts and Telecommunications, Beijing, China}
\IEEEauthorblockA{\IEEEauthorrefmark{2}Beijing University of Posts and Telecommunications Research Institute, Shenzhen, China
}
\IEEEauthorblockA{\IEEEauthorrefmark{3}Department of Computer Science, New York Institute of Technology
\\Email: \{zllin, houyanzhao\}@bupt.edu.cn}}

\maketitle
\begin{abstract}
In this paper, we propose a novel Adaptive Transmission Strategy to improve energy efficiency~(EE) in a wireless point-to-point transmission system with Quality of Service~(QoS) requirement, i.e., delay-outage probability considered. The transmitter is scheduled to use the channel that has better coefficients, and is forced into silent state when the channel suffers deep fading. The proposed strategy can be easily implemented by applying two-mode circuitry and is suitable for massive machine type communications (MTC) scenarios. In order to enhance EE, we formulate an EE maximization problem, which has a single optimum under a loose QoS constraint. We also show that the maximization problem can be solved efficiently by a binary search algorithm. Simulations demonstrate that our proposed strategy can obtain significant EE gains, hence confirming our theoretical analysis.
\end{abstract}


\IEEEpeerreviewmaketitle

\section{Introduction}
Massive machine type communications (MTC) constitute a typical scenario of Long Term Evolution (LTE) and 
the fifth-generation (5G) wireless networks\cite{1}. In this scenario, objects are equipped with sensors and microprocessors to collect information of surroundings, communicate with each other and execute smart applications. Applications of MTC span from health care systems, smart gird, emergency and disaster response to wireless sensor networks and the Internet of Things (IoT). In addition, one major benefit of MTC devices is that they can operate autonomously for years or even decades without human intervention. It is believed that more than 50 million MTC devices with various Quality of Service (QoS) requirements would be connected to communication networks by 2020\cite{2}, which brings a series of new challenges such as access congestion and unprecedented energy consumption problem. 

Considering massive MTC scenarios are characterized by a large number of connected devices operating for a long battery life, a 100-fold increase of energy efficiency~(EE) is expected in 5G networks as compared with that of 4G wireless networks~\cite{3}. EE refers to quantity of information bits per unit of energy consumption of the communication module (in bit/Joule). Therefore, a network with innovative energy saving schemes is in urgent need.

The issue of EE has aroused great interest in recent studies. Zappone and Jorswieck \cite{4} presented a unified EE analytical framework as well as a well-established measure of EE. They defined EE as the ratio of the system achievable data rate or effective capacity to the total power consumed, which is the sum of radiated power and static circuit power. Wu and Negi\cite{5} introduced the notion of effective capacity by relating the Shannon capacity to communication queuing delay in the form of a QoS metric. Furthermore, Liu and Wu\cite{6} derived the form of EE under a delay-outage probability constraint, by adopting the effective capacity model. Based on this, optimal power allocation strategies for the link-layer tradeoff between the effective capacity and EE in delay-limited point-to-point and MIMO wireless systems were developed and analyzed \cite{7,8,9}. Musavian, Le-Ngoc \cite{10} and \mbox{Cheng et al.} \cite{11} proposed power allocation schemes to maximize EE while guaranteeing the QoS of 5G wireless networks, under joint average and peak power limitations. Taking into account the empty buffer cases for the transmitter, 
Sinaie et al.\cite{12} and Xu et al.\cite{13} built up a more accurate EE analytical model. 

However, all the works above assume that the transmitter would transmit as long as backlog exists in the buffer, which is undesirable in consideration of energy reduction. Tolerable conditional transmission is more suitable for massive MTC scenarios than unconditional transmission \cite{6,7,8,9,10,11,12,13}, because massive MTC is characterized by transmitting a relatively low volume of non-delay-sensitive data \cite{3} and MTC devices are expected to be energy saving. Moreover, EE maximization problem could merely be solved with complicated approaches in extreme cases, e.g., under very loose QoS constraints \cite{10,11}. Thus, a low complexity method based on conditional transmission is necessary for massive MTC. 

Motivated by this observation, we develop an adaptive transmission strategy based on two-mode circuitry. Our aim is to exchange part of effective capacity for reduction in energy consumption by adapting the service process. In order to increase EE under delay-outage probability constraints, the transmitter seeks to transmit in good channel states and choose to remain silent when the system undergoes deep fading. Based on the strategy, we study the problem of EE maximization in a wireless point-to-point system while guaranteeing an upper-bounded target delay-outage probability. We firstly establish a QoS guaranteed system model and derive the analytical expressions of EE. Then we investigate the effects of the strategy on EE. For the model considered, a QoS exponent bound is obtained, which determines when these solutions indeed work. Furthermore, binary search algorithms are utilized to achieve a most energy-efficient schedule. Finally, we verify the benefit of the proposed strategy through simulations. The major contributions made in this paper are three-fold as summarized below:

i) We formulate a novel strategy based on two-mode circuitry to increase EE constrained with QoS requirements by adjusting the service process. This strategy possesses low complexity and outstanding suitability for massive MTC scenarios.

ii) We investigate the effectiveness of the strategy to improve EE, discuss the convexity conditions of the distribution function and further optimize the strategy to obtain the maximum EE by means of binary search algorithms.

iii) We conduct simulations to examine our strategy under different conditions. Simulation results, consistent with numerical results, demonstrate that our strategy can significantly enhance EE performance, especially for targeted loose QoS requirements and also prove the reliability of our algorithms. 

The remainder of this paper is organized as follows. Section~II describes the system model and the adaptive transmission strategy. In Section~III, the analytical expressions of EE together with EE optimization problem are elaborated. In Section~IV, algorithms are elicited to solve the EE optimization problem and numerical results as well as simulation results are presented to evaluate our strategy. Section~V summarizes our work.

\section{System Model}
\begin{figure}[!h]
\begin{center}
\includegraphics[width=7.6cm]{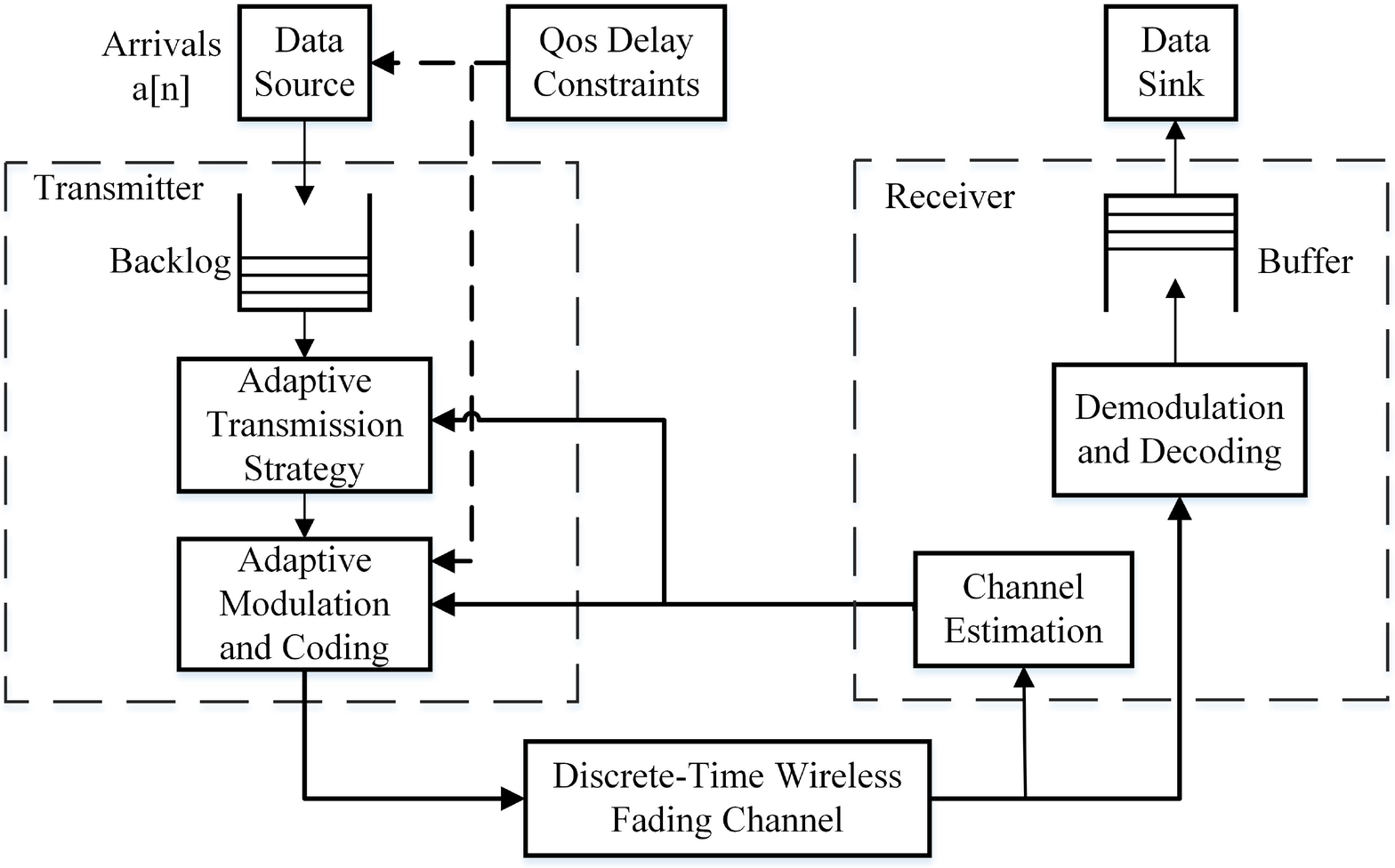}
\caption{A point-to-point system model with adaptive transmission strategy.}
\label{Fig.1}
\end{center}
\end{figure}
A wireless point-to-point system model over \mbox{Nakagami-$m$} block fading channels is considered, as shown in Fig.1. The system is time-discrete with time slot $T_s$. $T_s$ is assumed to equal the length of a fading block. Namely, the channel states maintain invariant during each time slot, but vary independently across time slots. As Fig.1 shows, the system consists of five basic components: data source, transmitter, wireless channel, receiver and data sink. 

Data arrives at the transmitter buffer from upper layers at a constant rate $\mu$. In time-slotted systems, $a[n]$ denotes the arrival data length in the $n$th slot $(n={1,2,3,...})$. Due to the constant arrival rate, $a[n]$ is identical across time slots and can be given as
\begin{equation}
a[n]=\mu{T_s}.
\end{equation}

The transmitter employs a buffer to backlog the data. Let $s[n]$ denote the actual service process under our adaptive transmission strategy in the $n$th slot. The queue length $q[n]$$(n={1,2,3,...})$, defined as the data length stored in the buffer at the end of time slot $n$, is updated as
\begin{equation}
q[n]=\max\{q[n-1]+a[n]-s[n],0\}.
\end{equation}
To restrict queuing delay, delay outage probability is limited by QoS exponent $\theta$ (see Section III.A for details).

The wireless channel is assumed to be a \mbox{Nakagami-$m$} block-fading channel. Define the channel service process $\tilde{s}[n]$$(n={1,2,3,...})$ as the amount of data the channel can support transmitting in the $n$th slot, which only depends on instantaneous channel capacity, regardless of our adaptive transmission strategy. Assume that $\tilde{s}[n]$ is stationary and ergodic. It is noted that $\tilde{s}[n]$ is different from $s[n]$; $s[n]$ has taken the adaptive transmission strategy into account by adapting $\tilde{s}[n]$ to the current channel state, thus  implying actual service process. 

According to the Shannon capacity formula, the service process $\tilde{s}[n]$ can be described as 
\begin{equation}
\tilde{s}[n]=T_sB_c{\log _2}(1 + \gamma[n]\frac{{P_{tr}}}{{{P_L}{N_0}{B_c}}}).
\end{equation}
where $P_{tr}$ is the transmission power, $P_L$ is the distance-based path loss, $N_0$ is noise spectral density, $B_c$ indicates the system bandwidth and $\gamma[n]$$(n={1,2,3,...})$ is the normalized channel power gain of the considered unit-variance Nakagami-$m$ block-fading channel with the probability density function~(PDF)\cite{14}:
\begin{equation}
f_\gamma(\gamma)=\frac{{{m^m}{{\gamma}^{m-1}}}}{{\Gamma(m)}}e^{-m\gamma}.
\end{equation}
where $\Gamma(\cdot)=\int_0^{\infty} w^{z-1}e^{-w}dw$ depicts the Gamma function and $m$ is the fading parameter.

The adaptive transmission strategy is formulated as follows: at the beginning of each slot, the adaptive transmission module determines whether or not to transmit data in the current time slot based on the collected information from the channel estimation module. It is assumed that the channel gain $\gamma[n]$ is perfectly known by the transmitter.

Set a normalized channel gain $\gamma[n]$ threshold as $\gamma_0$. $\gamma[n]\geq\gamma_0$ means the channel state is good and the transmitter transmits data to the receiver, which is defined as the transmission mode. Conversely, the system works in idle mode when $\gamma[n]<\gamma_0$, indicating no data transmission in the current slot. Thus, the service process $\tilde{s}[n]$ is adapted by the strategy and the actual service process can be expressed as
\begin{equation}
s[n]=\left\{
\begin{array}{rcl}
\tilde{s}[n],      &      & {\gamma[n]\geq\gamma_0},\\
0,     &      & {\gamma[n]<\gamma_0}.
\end{array} \right.
\end{equation}
\section{Energy Efficiency Analysis and Optimization}
In this section, general descriptions of effective capacity, delay-outage constraints and EE are introduced. On the basis of that, we focus on analyzing and optimizing EE with fixed system parameters.
\subsection{Effective Capacity and Delay-Outage Constraints}
According to \cite{5}, effective capacity represents the maximum arrival rate that a given service process can support subject to the QoS exponent $\theta$. For a block fading model, the log-moment generating function of service process can be written as
\begin{equation}
\Lambda^{(c)}(-\theta)={\ln}(\mathbb{E}\left[e^{-\theta{s[n]}}\right]),
\end{equation}
where $\mathbb{E}[\cdot]$ is an expectation operator. The QoS exponent~$\theta$ $(\theta>0)$ constrains the delay-outage probability. $\theta\to0$ indicates that the system has no limitation of delay, whereas $\theta\to\infty$ indicates that the system cannot tolerate any delay. Given an arrival process with constant rate, the queuing delay process\cite{15} converges in distribution to a random variable $D(\infty)$ such that
\begin{equation}
\lim_{D_{\max}\to\infty}
\frac{{\log\Pr\{D>D_{\max}\}}}{{D_{\max}}}
=-\theta,
\end{equation}
where $D_{\max}$ specifies a delay bound.

Suppose that\\
\indent 1) the assumptions of Gartner-Ellis theorem are satisfied,\\
\indent 2) $\Lambda^{(c)}(-\theta)$ is differentiable for all $\theta>0$.\\
Then the effective capacity is defined as
\begin{equation}
\begin{split}
\alpha^{(c)}(\theta)&=\frac{{-\Lambda^{(c)}(-\theta)}}{{\theta{T_s}}}\\
&=-\frac{{1}}{{\theta{T_s}}}
{\ln}(\mathbb{E}\left[e^{-\theta{s[n]}}\right]).
\end{split}
\end{equation}
And the probability that a delay exceeds the delay bound\cite{5} can be estimated as
\begin{equation}
\Pr\{Delay>D_{\max}\}\approx
{P_b}e^{-\theta^{(c)}(\mu){D_{\max}}},
\end{equation}
where $P_b$ is the probability that the transmitter buffer is nonempty at a randomly chosen slot $n$ and the QoS exponent $\theta^{(c)}(\mu)$ is defined as $\theta^{(c)}(\mu)=\mu{\alpha^{-1}(\mu)}$. $\alpha^{-1}(\mu)$ denotes the inverse function of $\alpha^{c}(\theta)$.

Based on our system model in Section II, by inserting (3) and (5) into (8), the effective capacity can be rewritten as
\begin{equation}
\begin{split}
\alpha^{c}(\theta)=&-\frac{{1}}{{\theta{T_s}}}
{\ln}\left(\int_0^{\gamma_0}f_{\gamma}(\gamma)d\gamma+\right.\\
&\left.\int_{\gamma_0}^{\infty}e^{-{\theta}T_sB_c\log_2\left(1+{\gamma} \frac{{P_{tr}}}{{P_LN_0B_c}}\right)}f_{\gamma}(\gamma)
d\gamma\!\!\right).
\end{split}
\end{equation}

\subsection{Definition of Energy Efficiency}
Aiming to achieve a tradeoff between effective capacity maximization and energy consumption reduction, EE is defined as the ratio of effective capacity to the total power consumption at the transmitter, namely
\begin{equation}
EE=\frac{{\alpha^{(c)}(\theta)}}{{P_{total}}}.
\end{equation}

For our system model with two-mode circuitry, $P_{total}$ can be expressed as
\begin{equation}
P_{total}=P_c+p_{tr}P_{tr}+p_{idle}P_{idle},
\end{equation}
where $P_c$ indicates circuit power consumption which is independent of data transmission, $P_{tr}$ and $P_{idle}$ respectively correspond to power consumption values for the transmission mode and the idle mode. It is reasonable to assume that $P_{tr}$ is larger than $P_{idle}$ because the transmission mode requires more radiated power. $p_{tr}$ and $p_{idle}$ respectively point out the probability of the system operating in the transmission mode and the idle mode. The sum of $p_{tr}$ and $p_{idle}$ equals one for two-mode circuitry.

Following our adaptive transmission strategy, $p_{tr}$ and $p_{idle}$ can be calculated by
\newcounter{mytempeqncnt2}
\begin{figure*}[!t]
\normalsize
\setcounter{mytempeqncnt2}{\value{equation}}
\setcounter{equation}{15}
\begin{small}
\begin{equation}
\label{eqn_dbl_x}
EE=\frac{{-\frac{{1}}{{\theta{T_s}}}
{\ln}\left(\int_0^{\gamma_0}\!f_{\gamma}(\gamma)d\gamma+\int_{\gamma_0}^{\infty}e^{-{\theta}T_sB_c\log_2\left(1+{\gamma} \frac{{P_{tr}}}{{P_LN_0B_c}}\right)}f_{\gamma}(\gamma)
d\gamma\right)}}
{{P_c+P_{tr}\int_{\gamma_0}^{\infty}f_{\gamma}(\gamma)d\gamma+P_{idle}\int_{0}^{\gamma_0}f_{\gamma}(\gamma)d\gamma}}.
\end{equation}
\end{small}
\setcounter{equation}{16}
\hrulefill
\vspace*{4pt}
\end{figure*}
\begin{figure*}[!t]
\normalsize
\setcounter{mytempeqncnt2}{\value{equation}}
\setcounter{equation}{16}
\begin{equation}
\begin{split}
\label{eqn_dbl_x}
EE&=\frac{{-\frac{{1}}{{\theta{T_s}}}{\ln}\left(1+e^{-2\gamma_0}(-2\gamma_0-1)+4\int_{\gamma_0}^{\infty}(1+m_2\gamma)^{m_1\theta}\gamma{e^{-2\gamma}}d\gamma\right)}}
{{P_c+P_{tr}e^{-2\gamma_0}(2\gamma_0+1)+P_{idle}\left(1+e^{-2\gamma_0}(-2\gamma_0-1)\right)}}\\
&=\frac{{-\frac{{1}}{{\theta{T_s}}}{\ln}\left(1+e^{-2\gamma_0}(-2\gamma_0-1)+2^{-m_1\theta}{m_2}^{m_1\theta}\Gamma(2+m_1\theta,2\gamma_0)\right)}}
{{P_c+P_{idle}+(P_{tr}-P_{idle})e^{-2\gamma_0}(2\gamma_0+1)}}.
\end{split}
\end{equation}
\setcounter{equation}{17}
\hrulefill
\vspace*{4pt}
\end{figure*}
\begin{figure*}[!t]
\normalsize
\setcounter{mytempeqncnt2}{\value{equation}}
\setcounter{equation}{17}
\begin{small}
\begin{equation}
\label{eqn_dbl_x}
\begin{split}
\frac{{\partial{EE}}}{{\partial\gamma_0}}=&-\frac{{4\gamma_0e^{-2\gamma_0}}}
{{\theta{T_s}\left(P_c+P_{idle}+(P_{tr}-P_{idle})e^{-2\gamma_0}(2\gamma_0+1)\right)^{2}F(\gamma_0)}} \\
&\Bigg(\left(1-(1+m_2\gamma_0)^{m_1\theta}\right)
\left(\!P_c+P_{idle}+(P_{tr}-P_{idle})e^{-2\gamma_0}(2\gamma_0+1)\right)
+(P_{tr}-P_{idle}){\ln}\left(F(\gamma_0)\right)F(\gamma_0)\Bigg).
\end{split}
\end{equation}
\end{small}
\setcounter{equation}{12}
\hrulefill
\vspace*{4pt}
\end{figure*}

\begin{equation}
p_{tr}=\int_{\gamma_0}^{\infty}f_{\gamma}(\gamma)d\gamma,
\end{equation}
\begin{equation}
p_{idle}=\int_{0}^{\gamma_0}f_{\gamma}(\gamma)d\gamma.
\end{equation}
Hence, the total power consumption is obtained as
\begin{equation}
P_{total}=P_c+P_{tr}\int_{\gamma_0}^{\infty}f_{\gamma}(\gamma)d\gamma+P_{idle}\int_{0}^{\gamma_0}f_{\gamma}(\gamma)d\gamma.
\end{equation}

By inserting (10) and (15) into (11), we have (16) as shown at the top of this page.

\subsection{Optimization of Energy Efficiency}
In this section, we analyze the strategy's impact on EE. The Nakagami-$m$ fading parameter $m=2$ (i.e., approximate Rice fading)\cite{16}. Then EE could be rewritten as (17) at the top of this page, where $m_1=-T_sB_c\log_2e$, $m_2=\frac{{P_{tr}}}{{P_{L}N_0B_c}}$ for convenience and $\Gamma(v,z)=\int_z^{\infty}w^{v-1}e^{-w}dw$ denotes the upper incomplete Gamma function.

In order to optimize $\gamma_0$ with variable QoS exponent~$\theta$, we compute the first-order derivative of (17), which is given by (18) at the top of this page, where
\setcounter{equation}{18}
\begin{equation}
\begin{split}
F(\gamma_0)&=1+e^{-2\gamma_0}(-2\gamma_0-1)\\
&+2^{-m_1\theta}{m_2}^{m_1\theta}\Gamma(2+m_1\theta,2\gamma_0).
\end{split}
\end{equation}
By calculating the first derivative of $F(\gamma_0)$, it can be found that $F(\gamma_0)$ is an increasing function of $\gamma_0$. Hence, we have
\begin{equation}
\begin{split}
F(\gamma_0)>&F(\gamma_0=0)\\
&=2^{-m_1\theta}{m_2}^{m_1\theta}\Gamma(2+m_1\theta,2\gamma_0)\\&>0
\end{split}
\end{equation}
and
\begin{equation}
F(\gamma_0)<F(\gamma_0=\infty)=1.
\end{equation}
Thus, $F(\gamma_0)$ is bounded that 
\begin{equation}
0<F(\gamma_0)<1\ (\theta>0).
\end{equation}
Concerning that 
\begin{equation}\nonumber
\frac{{4\gamma_0e^{-2\gamma_0}}}
{{\theta{T_s}\!\left(P_c\!\!+\!\!P_{idle}\!\!+\!\!(P_{tr}\!\!-\!\!P_{idle})e^{-\!2\gamma_0}(2\gamma_0\!\!+\!\!1)\right)^{\!2}F(\!\gamma_0)}}\!\!>\!\!0
\end{equation}
holds for all $\theta>0$ and $\gamma_0>0$, the positive or negative of $\frac{{\partial {EE}}}{{\partial\gamma_0}}$ is identical to
\begin{equation}
\begin{split}
&\!G(\!\gamma_0\!)\!=-(P_{tr}\!-\!P_{idle}){\ln}\left(F(\gamma_0)\right)F(\gamma_0)\\
&-\!\!\left(\!1\!\!-\!\!(\!1\!+\!m_2\gamma_0)^{\!m_1\!\theta\!}\right)\!\!\left(\!P_c\!\!+\!\!P_{idle}\!\!+\!\!(\!P_{tr}\!\!-\!\!P_{idle}\!)
(2\gamma_0\!\!+\!\!1)e^{-\!2\gamma_0}\!\right)\!,
\end{split}
\end{equation}
which can be utilized to analyze the increasing or decreasing trend of EE.

Based on (22) and that $\!P_{tr}\!$ is larger than $\!P_{idle}$, $\!\!-\left(\!P_{tr}\!\!-\!\!P_{idle}\!\right){\ln}\!\left(F(\!\gamma_0)\right)F(\gamma_0)$\ \!is sure to be positive while
\begin{equation}\nonumber
\begin{small}
 \!-\!\!\left(\!P_c\!\!+\!\!P_{idle}\!\!+\!\!(\!P_{tr}\!\!-\!\!P_{idle}\!)e^{\!\!-\!2\gamma_0}(2\gamma_0\!\!+\!\!1\!)\right)\!\!\left(\!1\!\!-\!\!(1\!+\!m_2\gamma_0\!)^{\!m_1\!\theta}\right)
\end{small} \end{equation}
is negative. Hence, the trend of $G(\gamma_0)$ is variable, largely dependent on $\theta$.


\begin{table}[h]
\centering
\caption{System Parameters}
\renewcommand\arraystretch{1.17}
\begin{tabular}{|l|l|}
\hline
\textbf{Parameters} & \textbf{Values} \\
\hline
Time duration of a slot, $T_s$ & 1ms  \\
\hline
Distance, $d$ & 1km \\
\hline
Noise spectral density, $N_0$ & -174dBm/Hz \\
\hline
System Bandwidth, $B_c$ & 180kHz \\
\hline
Fading parameter, $m$ & 2 \\
\hline
Circuit power, ${P_{tr}}$ & 43dBm \\
\hline
Transmission power, ${P_c}$ & 0.1watt \\ 
\hline
Idle mode power, ${P_{idle}}$ & 0 \\ 
\hline
\end{tabular}
\end{table}
\begin{figure}[!htb]
\begin{center}
\includegraphics[width=8cm]{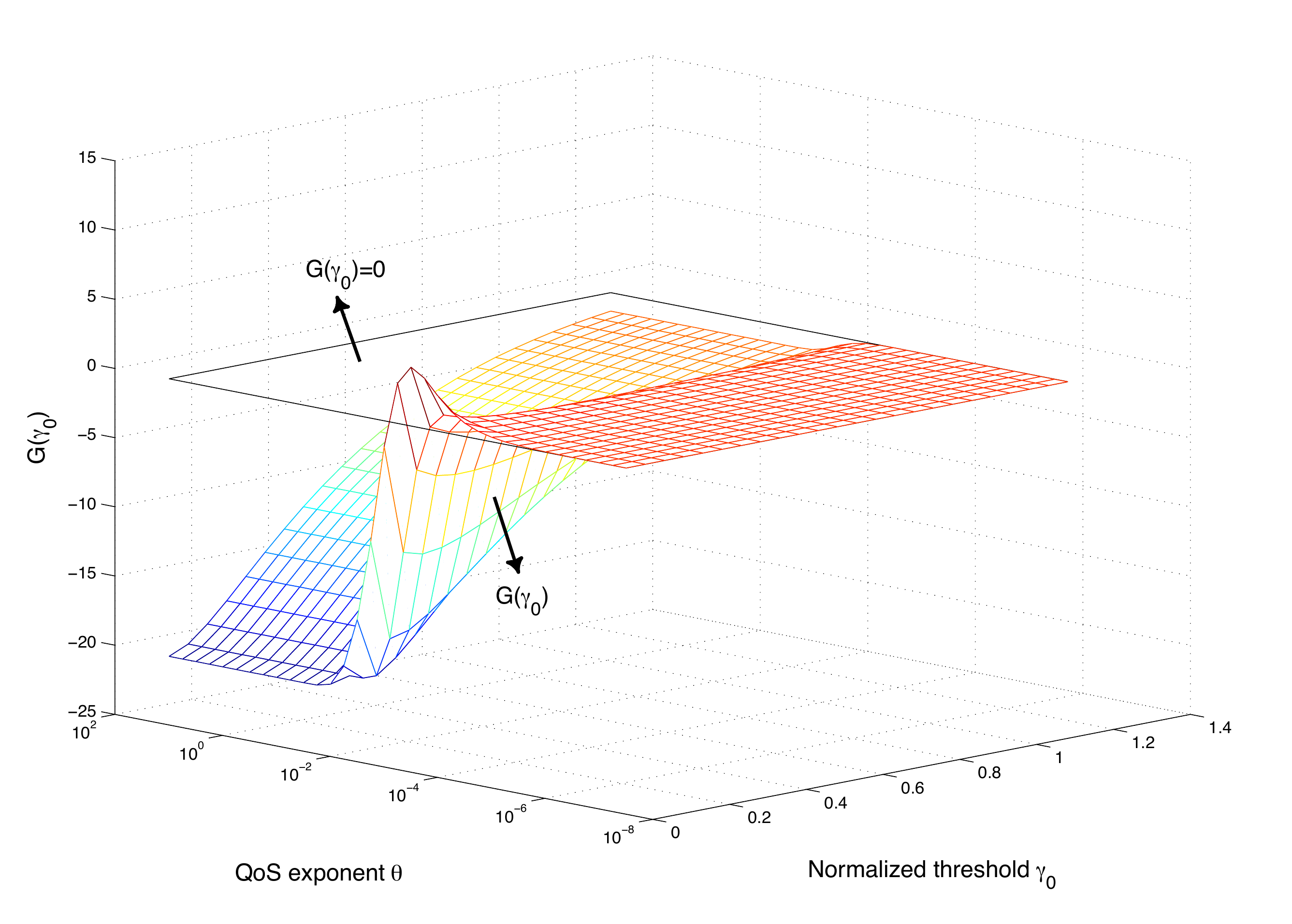}
\caption{Numerical results of $G(\gamma_0)$ versus normalized threshold~$\gamma_0$ and QoS exponent~$\theta$ and a reference plane $G(\gamma_0)=0$.}
\label{Fig.2}
\end{center}
\end{figure}
Considering that the closed-form solution for \mbox{$\frac{{\partial {EE}}}{{\partial\gamma_0}}\!=\!0$} is hard to derive, we employ numerical methods to further analyze EE. The system parameters are depicted in Table I and the QoS exponent $\theta$ is a variable to be determined by user requirements in terms of delay. Referring to \cite{17}, the distance-based path loss $P_L$ in (3) can be calculated by 
\begin{equation}P_{L}({\rm{dB}})=128.1+37.6\log_{10}(d),
\end{equation}
where $d$ illustrates the distance between the transmitter and the receiver and is specified as 1km in consideration of enhanced coverage and low mobility features of the MTC scenario.

By following the system settings in Table I, Fig.2 plots 
$G(\gamma_0)$ and the plane $G(\gamma_0)=0$ as reference. It characterizes $G(\gamma_0)$ versus the QoS exponent $\theta$ and the normalized threshold $\gamma_0$. As Fig.2 shows, a QoS exponent threshold $\theta_{thr}$ exists for the current system model that for all $\theta<\theta_{thr}$, $G(\gamma_0)$ would be positive for relatively small $\gamma_0$ while for $\theta\geq{\theta_{thr}}$, $G(\gamma_0)$ is always negative, which is summarized as two cases below.

\emph{CASE I:} In the case of $\theta<{\theta_{thr}}$, an optimal $\tilde{\gamma_0}$ exists that $G(\gamma_0)>0$ for $0<\gamma_0<\tilde{\gamma_0}$ and $G(\gamma_0)<0$ for $\gamma_0>\tilde{\gamma_0}$. As a positive or negative $G(\gamma_0)$ corresponds to increasing or decreasing trend of EE, EE turns out to be a unimodal function. It is concluded that EE can be maximized if the optimal $\tilde{\gamma_0}$ is appropriately found. The corresponding algorithms and results are presented in Section~IV.

\emph{CASE II:} In the case of $\theta\geq{\theta_{thr}}$, $G(\gamma_0)$ is undoubtedly negative for all $\gamma_0>0$. Therefore, EE monotonically decreases as $\gamma_0$ increases, implying that it is unnecessary to bring in the strategy.

\section{Algorithms and Results}
\begin{figure}[!htb]
\begin{center}
\includegraphics[width=8cm]{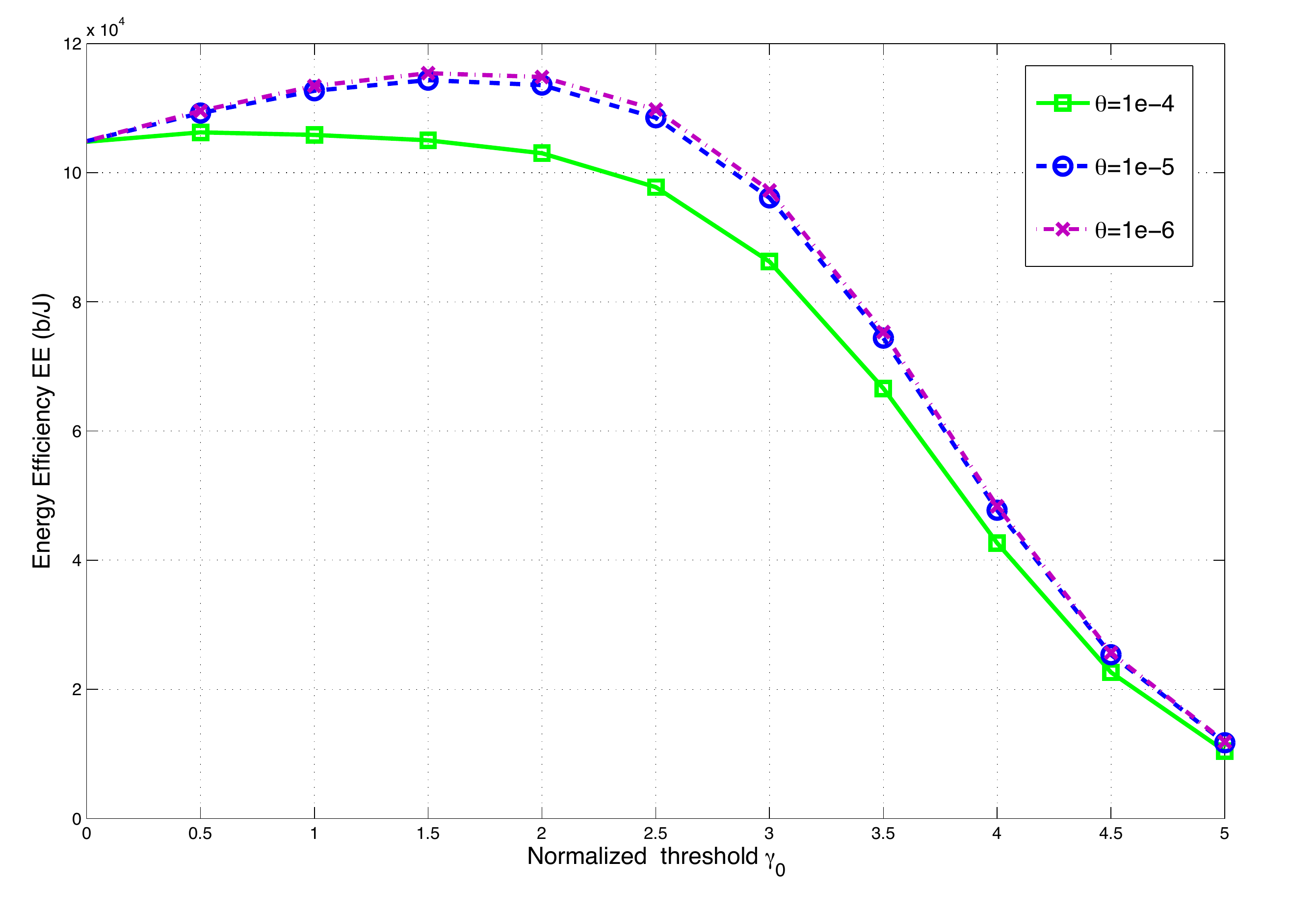}
\caption{Numerical results of EE versus normalized threshold $\gamma_0$ for various QoS exponent $\theta$.}
\label{Fig.3}
\end{center}
\end{figure}
With the system parameters settled as in Table I, Fig.3 plots the numerical EE versus normalized threshold $\gamma_0$ according to (17) with QoS exponent $\theta$ set to  $10^{-4}$, $10^{-5}$ and $10^{-6}$, respectively. As an example, $\theta=10^{-4}$ indicates that the delay-outage probability is required to be less than 0.1 when assuming the delay bound as 10 ms,  according to (9). Fig.3 confirms the existence of EE convex point and the optimal $\tilde{\gamma_0}$ when QoS exponent $\theta$ is smaller than $10^{-4}$. It should be pointed out that the special case $\gamma_0=0$ signifies that the transmitter transmits irrespective of channel state, implying that the strategy is not in use. Thus, EE$(\gamma_0=0)$ equals EE under traditional system model without the adaptive transmission strategy.

As Fig.3 shows, the proposed strategy could effectively enhance EE compared to the condition without the strategy (EE$(\gamma_0=0)$). Particularly, a smaller $\theta$ setting (a looser delay requirement) would lead to a larger optimal $\tilde{\gamma_0}$ value and also bring in more delightful increase in EE. That is because with a smaller $\theta$, a longer delay could be tolerated by the user, which leaves more available space of effective capacity to be sacrificed to improve EE.

\begin{algorithm}[h] 
\caption{ Optimal $\tilde{\gamma_0}$ Search Algorithm} \label{alg:Algorithm 1} 
\begin{algorithmic} 
\STATE 1: Initialize\\
\ \ \ $\varepsilon$  \{a precision tolerance, e.g.,$\varepsilon=10^{-8}$\},\\
\ \ \ $\gamma_0^l$  \{a lower bound, e.g.,$\gamma_0^l=0$\},\\
\ \ \ $\gamma_0^u$  \{an upper bound, guaranteeing $G(\gamma_0^u)<0$\};
\STATE 2: $\varepsilon_t=\gamma_0^u-\gamma_0^l$;
\STATE 3: $\gamma_0^t=(\gamma_0^u+\gamma_0^l)/2$;
\STATE 4: \textbf{while} $|\varepsilon_t|>\varepsilon$  \textbf{do} 
\STATE 5: \ \ \ \textbf{if} $G(\gamma_0^t)<0$\ \textbf{then} 
\STATE 6: \ \ \ \ \ $\gamma_0^u=\gamma_0^t$;
\STATE 7: \ \ \ \textbf{else} 
\STATE 8: \ \ \ \ \ $\gamma_0^l=\gamma_0^t$;
\STATE 9: \ \ \ \textbf{end if} 
\STATE 10:\ \ \ $\varepsilon_t=\gamma_0^u-\gamma_0^l$;
\STATE 11:\ \ \ $\gamma_0^t=(\gamma_0^u+\gamma_0^l)/2$;
\STATE 12: \textbf{end while}
\STATE 13: $\tilde{\gamma_0}=\gamma_0^t$;
\end{algorithmic}
\end{algorithm}

With these observations, binary search algorithm\cite{18} is proposed to determine the unique QoS exponent threshold $\theta_{thr}$ and the optimal $\tilde{\gamma_0}$. The optimal $\tilde{\gamma_0}$ search algorithm is described in Algorithm 1. Setting the system parameters as in Table I, we find the value of $\theta_{thr}$ rounded to $7.219\times10^{-4}$ via the algorithm with a precision tolerance $\varepsilon$ being $10^{-8}$. Based on analysis in Section III.C, a higher EE is achievable through our strategy under loose delay constraints (i.e., $\theta<7.219\times10^{-4}$), which matches well with Fig.3.

\begin{table}[htbp]
\centering
\caption{Numerical Results}
\renewcommand\arraystretch{1.17}
\begin{tabular}{ p{1.05cm}<{\centering} p{1.3cm}<{\centering} p{1.9cm}<{\centering} c}
\hline
\tabincell{c}{\textbf{QoS} \\ \textbf{exponent}\\ \textbf{$\theta$}}  
&\tabincell{c}{\textbf{Optimal}\\\textbf{normalized} \\\textbf{threshold $\tilde{\gamma_0}$} }
&\tabincell{c}{\textbf{Maximum EE}\\ \textbf{with the strategy}\\ \textbf{(bits/J)} }
&\tabincell{c}{\textbf{EE without}\\ \textbf{the strategy}\\ \textbf{(bits/J)} }\\
\hline
$1\times10^{-4}$& 0.5323 & $1.0623\times10^{5}$& $1.0478\times10^{5}$\\
\hline
$1\times10^{-5}$& 1.6293 & $1.1441\times10^{5}$& $1.0488\times10^{5}$\\
\hline
$1\times10^{-6}$& 1.6606 & $1.1544\times10^{5}$& $1.0489\times10^{5}$\\
\hline
$1\times10^{-7}$& 1.6636 & $1.1554\times10^{5}$& $1.0489\times10^{5}$\\
\hline
\end{tabular}
\end{table}

Table II lists a slice of numerical results of $\tilde{\gamma_0}$ derived via the algorithm with a precision tolerance $\varepsilon$ being $10^{-8}$. $\theta$ is set as $10^{-4}$, $10^{-5}$, $10^{-6}$ and $10^{-7}$, respectively. In addition, the corresponding maximum EE with and without the adaptive transmission strategy are also presented according to (17). These results are consistent with Fig.3. Hence, the binary search algorithm is proved solid. Especially, given the QoS exponent $\theta=10^{-4}$, we achieve the optimal $\tilde{\gamma_0}$ of approximately 0.5323 with a precision tolerance $\varepsilon$ being $10^{-8}$. The corresponding effective capacity $\alpha^{(c)}(\theta)$ is obtained as 1519.7 kbps and the corresponding EE equals $1.062\times10^5$ bits/J. 


The system model in Section II is simulated with parameters listed in Table I under different data arrival rates $\mu$. The QoS exponent $\theta$ is set as $10^{-4}$ and the arrival rates $\mu$ are 300 kbps and 1519.7 kbps, respectively. Each simulation runs 200,000 time slots.

Fig.4 plots the simulated EE results, denoted as the ratio of data arrival rate to the total power consumption at the transmitter, and numerical results as reference. Considering both the QoS exponent $\theta$ and the arrival rate $\mu$ are determined, there is an upper bound of threshold $\gamma_0$ in case that the arrival rate exceeds the effective capacity. According to (10), the upper bound of threshold $\gamma_0$ can be approximately computed, being 0.53 for 1519.7 kbps and 1.73 for 300 kbps. 

\begin{figure}[!h]
\begin{center}
\includegraphics[width=8cm]{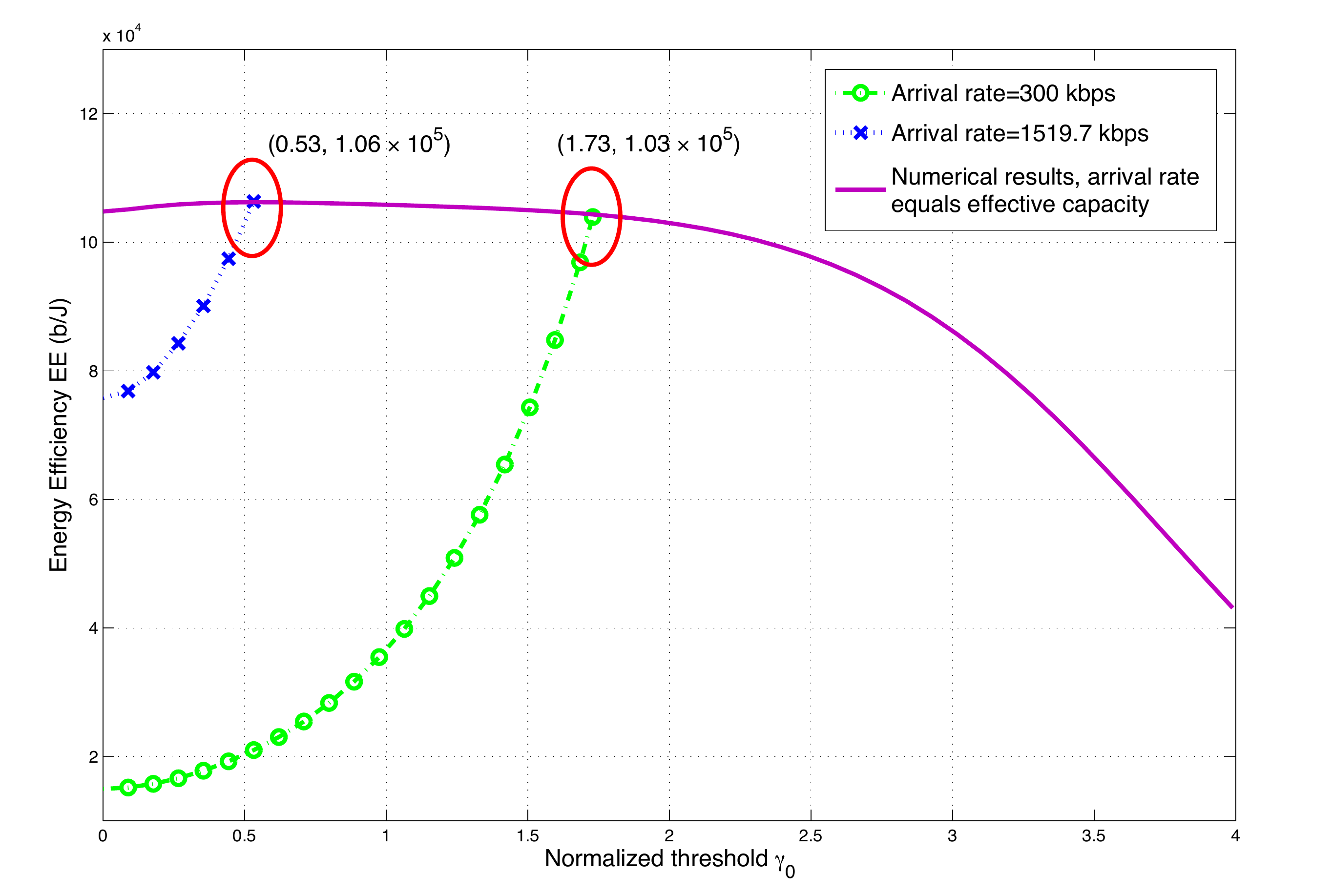}
\caption{Simulation results of EE versus normalized threshold $\gamma_0$ with data arrival rate $\mu=$300 kbps, 1519.7 kbps and numerical results, both restricted with QoS exponent $\theta=10^{-4}$. The simulation curves intersect the numerical curve at points  ($0.53$, $1.06\times10^5$) and ($1.73$, $1.03\times10^5$), marked with red circles.}
\label{Fig.4}
\end{center}
\end{figure}
The two simulated EE curves in Fig.4 (labeled arrival rate 300~kbps and 1519.7kbps) depict the impact of our strategy on EE under constant data arrival rate. We can see that EE increases with the threshold $\gamma_0$ until $\gamma_0$ reaches the upper bound. Fig.4 shows 39.86\% and 589.56\% EE improvement compared to the condition without the strategy (EE$(\gamma_0=0)$) for the arrival rate $\mu$ given as 1519.7 kbps and 300 kbps, respectively. This improvement is attributed to the introduction of two-mode circuitry, which saves energy for the transmitter by exploiting the time varying nature of the wireless channel and spending more time in idle mode.

In addition, Fig.4 reveals that the simulated EE curves, with arrival rate 1519.7 kbps and 300 kbps, intersect the numerical EE curve respectively at $\gamma_0=0.53$ and $\gamma_0=1.73$ (their upper $\gamma_0$ bounds). Note that the numerical results are derived with the assumption that data arrival rate is a variable equal to effective capacity. Thus, at the intersection points, these two arrival rates have reached the effective capacity of the numerical EE curve. Moreover, the values of threshold $\gamma_0$, arrival rate as well as EE at the intersection point ($0.53$, $1.06\times10^5$) match well with the results given by our algorithms. Therefore, Fig.4 shows that the simulation results are consistent with numerical results and that our algorithms are reliable. The effectiveness of our strategy in terms of improving EE is hence proved. 

Furthermore, numerical and simulation results demonstrate that our adaptive transmission strategy is favorable for massive MTC scenarios. In the first place, in massive MTC scenarios, tremendous number of devices may transmit only occasionally, at low bit rate\cite{3}. And beneficially, adopting the strategy, a lower bit rate leads to more EE enhancements, as shown in Fig.4. Secondly, MTC scenarios are of various delay requirements, which could be satisfied by adapting the strategy. And the strategy would bring in more EE improvements in non-delay-sensitive MTC scenarios such as smart grid and smart agricultural system.

\section{Conclusion}
In this paper, we develop an adaptive transmission strategy under delay-outage probability guarantees. The scenario discussed in this work is a typical point-to-point wireless communication system with constant data arrival rate. Based on theoretical analysis, a parameter involved in the strategy is optimized to maximize EE. Numerical results confirm the existence of EE convex points under loose QoS requirements. Furthermore, both simulation results and numerical results indicate that EE could attain significant enhancements with our strategy. Further extension of this study would be to refine the adaptive transmission strategy with threshold gradients and take a multi-user communication system into consideration.

\section*{Acknowledgment}
This work was supported in part by the National Science and Technology Major Project of China under Grant 2017ZX03001004, the National Nature Science Foundation of China Project (Grant No.61701042), the Shenzhen Science and Technology Project (No. JSGG20150512153045135) and 111 Project of China under Grant B16006.





%

\end{document}